Tuning the polarization-induced free hole density in nanowires graded from GaN to AlN

A. T. M. Golam Sarwar[1], Santino D. Carnevale[1], Thomas F. Kent[2], Fan Yang[2], David W. McComb[2], and Roberto C. Myers[1,2*].

[1]Department of Electrical and Computer Engineering, The Ohio State University, Columbus, Ohio 43210, USA.

[2]Department of Materials Science and Engineering, The Ohio State University, Columbus, Ohio 43210, USA

*E-mail: myers.1079@osu.edu.

Abstract

We report a systematic study of p-type polarization induced doping in graded AlGaN nanowire light emitting diodes grown on silicon wafers by plasma-assisted molecular beam epitaxy. The composition gradient in the p-type base is varied in a set of samples from 0.7 %Al/nm to 4.95 %Al/nm corresponding to negative bound polarization charge densities of $2.2 \times 10^{18}$ cm$^{-3}$ to $1.6 \times 10^{19}$ cm$^{-3}$. Capacitance measurements and energy band modeling reveal that for gradients greater than or equal to 1.30 %Al/nm, the deep donor concentration is negligible and free hole concentrations roughly equal to the bound polarization charge density are achieved up to $1.6 \times 10^{19}$ cm$^{-3}$ at a gradient of 4.95 %Al/nm. Accurate grading lengths in the p- and n-side of the pn-junction is extracted from scanning transmission electron microscopy images and is used to support energy band calculation and capacitance modeling. These results demonstrate the robust nature of p-type polarization doping in nanowires and put an upper bound on the magnitude of deep donor compensation.

Due to the lack of centrosymmetry in the wurtzite crystal structure, III-N compound semiconductors (i.e. AlN, GaN, and InN) and their alloys exhibit spontaneous and piezoelectric polarization [1] leading to net charge at surfaces and interfaces useful for a variety of devices [2] [3] [4] [5] [6]. A graded heterojunction of AlGaN leaves a three-dimensional (3D) slab of bound charge which can induce either an n- [7] or a p-type conducting layer [8] depending on the sign of the Al concentration gradient. A positive %Al/nm gradient along [0001] (III-face) generates a positive bound charge which is compensated by electrons that diffuse inwards from surface donor states. Vice versa, a negative %Al/nm along [0001] (identical to a positive %Al/nm along N-face [000$\bar{1}$]) generates a negative bound charge which can be compensated by free holes. Holes may be supplied by Mg-acceptors, or Be [9]. Grading over the full composition range in thin films (i.e. GaN to AlN) is limited by the small critical thickness for epitaxial strain relaxation. However, in nanowires the critical thickness increases as the nanowire diameter decreases, becoming infinite below a certain diameter [10]. Taking advantage of this strain accommodation, Carnevale et al fabricated composition graded AlGaN nanowires across the full composition range in a back and forth fashion leading to alternating p- and n-type regions [11]. This impurity-free pn diode [11] was later demonstrated in graded AlGaN thin films [12]. Although polarization-induced hole doping in graded thin films and nanowires is conceptually established, the origin of the free holes and the degree to which native donor compensation limits polarization-induced hole doping is not yet reported.

Here we perform a rigorous study of p-type polarization induced doping in AlGaN nanowires as a function of the magnitude of the Al concentration gradient. We systematically increase the negative bound polarization charge in the p-type base of nanowire LED structures by more steeply grading the composition. Capacitance measurements together with energy band modeling reveal that increasing the negative bound polarization charge proportionately increases the free hole density in the p-type section.

The nanowire diode design used in this study is shown in figure 1(a) and is similar to doubly graded polarization-induced AlGaN nanowire LEDs [11]. However, the nanowires are integrated with p-type Si (111) substrates to take advantage of the dominant N-face [000$\bar{1}$] polarity of nanowires [13]. The base of the nanowire heterostructures is defined by a linear compositional grading from GaN to AlN along [000$\bar{1}$] inducing a negative spontaneous polarization charge balanced by free holes, thereby inducing p-type conductivity. The valence band becomes flat and the bandgap gradient results in a sloped conduction band edge. An active region consisting of 3 periods of 5 nm GaN quantum wells (QW) and 5 nm AlN barriers (BR) is defined. The top of the structure is linearly graded from AlN to GaN, which creates positive polarization induced bound charge resulting in an n-type layer. Considering a linear grading profile in the nanowires, the spontaneous polarization induced bound charge is $N_A^{pol} = (P_{AlN} - P_{GaN})/L_{grad}$, where $P_{AlN}$ and $P_{GaN}$ are spontaneous polarization of AlN and GaN, respectively, and $L_{grad}$ is the length of the graded layer. To include the effect of piezoelectric polarization, we carry out 3D strain modeling using NextNano3 [14] (Fig. 1b). Nanowire heterostructures are expected to be strain-relaxed due to the high surface to volume ratio. GaN/AlN has a lattice mismatch of 2.41% and for very short grading length this nanowire heterostructure might be partially strained. For very large strain mismatch in InN to GaN graded nanowires (11%) cracking can occur [15]. However, according to Glas [10], the critical thickness for 2.4% lattice mismatch diverges at diameters less than 140 nm, and all nanowires in our study exhibit diameters less than 50 nm. The 3D strain simulation also verifies this theoretical prediction. Apart from the active region, most of the graded nanowire heterostructure is strain relaxed. Maximum strain is observed along the axis of the nanowire heterostructure in the 1$^{st}$ and 3$^{rd}$ GaN QWs ($\varepsilon_{xx}$ = 1.62%). Strain reduces away from the active region and falls below 0.1% within ~9 nm. The strain figures are even lower away from the axis of the nanowire. Figure 1(c) shows the calculated energy band diagram and carrier concentration profile along the axis of the nanowire structure of figure 1(a) considering the effect of both spontaneous and strain induced (piezoelectric) polarization charge [6]. Both the base and top grading length are taken as 50 nm with the Al concentration varying across the full range from GaN to AlN. The energy band diagram is simulated using technology computer aided design software ATLAS. A ~2eV valence band barrier is present at the p-Si/p-GaN interface due to the difference in the electron affinities.

Self-assembled, catalyst free GaN/AlN nanowire heterostructures are grown by plasma assisted molecular beam epitaxy (PAMBE) along [000$\bar{1}$] [13]. Using a two-step growth process [16], Mg-doped GaN nanowires are first nucleated on p-Si at 720C for 5 minutes. The substrate temperature is then set to higher temperature (790C) with a ramp rate of 25C. Composition grading is realized at high temperature (790C) using shutter pulsing [15] of Ga and Al to define the concentration within each monolayer. A set of five samples is prepared in which the length of the p-type base is varied from 25 to 200 nm, while the rest of the heterostructure is kept identical. To boost the conductivity of the graded regions, p-type base and n-type top are doped with Mg and Si, respectively.

Capacitance is one of the most well developed methods to measure the free carrier density in semiconductors. Here we develop a simple one-dimensional (1D) circuit model to predict the capacitance of the nanowire heterostructure. The capacitance of a semiconductor heterostructure is determined by the depletion regions. Shown in figure 1(c), there are two depletion regions in the heterostucture: at the base of the nanowire (W$_b$) and in the pn junction (Wj = W$_n$ + W$_{qw}$ +

$W_p$), which is further divided into three depletion sub-regions: n-side ($W_n$), active region ($W_{qw}$), and p-side ($W_p$). We therefore consider two capacitors in series: capacitance due to depletion at the base of the nanowire ($C_b = \epsilon/W_b$), and p-n junction capacitance ($C_j = \epsilon/(W_p+W_{qw}+W_n)$) around the active region of LED. Here, $\epsilon = \epsilon_r\epsilon_0$. $\epsilon_r$ is the relative dielectric constant and $\epsilon_0$ is the permittivity of free space. For capacitance calculations we use $\epsilon_r = 9$. The definition of $W_b$, $W_p$, $W_{qw}$, $W_n$ along with the corresponding energy band diagram and the schematic of the 1D circuit model are shown in figure 2(a).

High angle annular dark field (HAADF) scanning transmission electron microscopy (STEM) imaging (figure 2(b)) are used to extract the exact length of p-type (bottom) and n-type (top) sections of different samples in order to accurately model the energy band diagram and depletion widths. STEM sample were prepared with a Helios NanoLab 600 dual-beam focused ion beam (FIB). High resolution STEM imaging were carried out on an FEI image corrected Titan3TM G2 60-300 S/TEM at 300 kV. We find the average length of the n-type top section is 56.45±2.50 nm. GaN quantum well and AlN barrier thickness from the high resolution STEM (not shown) are found to be 2.70 ~ 3.0 nm. To find the accurate length of the p-type graded section in the grown samples we plotted, shown in figure 2 (c), the base length from the STEM images (figure 2(b)) versus the target (nominal) graded base length. A linear fit of the data points shows a y-intercept at 16.8 nm corresponding to the height of the GaN base grown during the nanowire nucleation step. This length (16.8 nm) is subtracted from the STEM extracted base length to find the exact p-type grading length (GaN to AlN) from which we can calculate the negative bound charge density due to spontaneous polarization. Nominal graded length ($L_{nom}$), Achieved grading length ($L_{grad}$), corresponding polarization gradient (%Al/nm), and spontaneous polarization charge ($N_A^{pol}$) are listed in table I. The piezoelectric charge is calculated from the 3D strain simulation.

The 1D energy band diagram considering the effect of spontaneous and strain-induced polarization charge of each of the devices is calculated using the revised dimensions from STEM images to find $W_b$, $W_p$, $W_{qw}$, and $W_n$. The edge of the depletion region is defined where the mobile carrier (electrons/holes) concentration drops below $10^{18}$ cm$^{-3}$ (dashed blue line in figure 1(c)). Total 1D capacitance ($C_{1D}$) is calculated using $C_{1D} = (1/C_b + 1/C_j)^{-1}$. Triangle symbols in figure 3 represent the calculated capacitance using the described 1D model.

Capacitance - voltage (C-V) measurements are performed on nanowire diodes fabricated using a top 10 nm Ti/ 20 nm Au contact. Nanowires are mechanically removed from the Si wafer and In is soldered to form a diffused bottom contact. We note here that unlike planar structures nanowire devices have a less than one fill factor and the effect of fill factor must be accurately included. The planar fill factor ($ff_p$) is measured from plan view scanning electron microscopy (SEM) images using ImageJ software. Shown in fig. 2(d), the visible nanowire tops are converted into black color while rest of the area is kept white. The $ff_p$ is then calculated from the ratio of black to total area. From cross sectional SEM images, we measure the degree of nanowire taper and quantify it from of the base to top area ratio ($r_{bt} = (d_{base}/d_{top})^2$), and the QW to top area ratio ($r_{qt} = (d_{qw}/d_{top})^2$) of the nanowires (see in fig. 2(e)). Table II lists all the values of $ff_p$, $r_{bt}$, and $r_{qt}$ for all the nanowire devices under study. The total nanowire capacitance is calculated from $C_{cyl} = 1/(C_b^{-1}+C_j^{-1})$, assuming a cylindrical shape. Tapering leads to an additional scaling that alters the total nanowire capacitance to $C_{tap} = 1/\{(r_{bt}C_b)^{-1}+(r_{qt}C_j)^{-1}\}$. The fill factor is also modified, considering the tapered geometry, $ff_{tap} = C_{tap}/C_{cyl}$. The total effective fill factor incorporates the actual planar fill factor and effect of the tapering, $ff_{eff} = ff_p\, ff_{tap}$. Table III lists all the calculated values for ff$_{tap}$, and ff$_{eff}$. C$_b$ and C$_j$ in table III are calculated from the energy band diagram and charge concentration as described previously.

We extract the equivalent 1D capacitances of the graded base nanowire LEDs using the total effective fill factor and measured capacitance using

$$C_{measured} = C_{air} + C_{nw} = (1 - ff_{eff})\frac{\varepsilon_0}{h} + ff_{eff}\, C_{1D\_extracted},$$

where $\varepsilon_0$ is free space permittivity and $h$ is average nanowire height. Square symbols in figure 3 represent the extracted 1D capacitance measured for the polarization graded nanowire diodes. The error bars include the uncertainty in effective fill factor calculation based on the uncertainty and statistics of the SEM based nanowire measurements. For comparison, we also model the 1D capacitance assuming the p-type base is fully depleted of holes (green circles in figure 3). For short base lengths (< 80 nm) the extracted capacitance of nanowire devices agrees with the 1D capacitance model, while far off from the fully depleted model. For base lengths greater than 80 nm measured capacitances are in between the two models. This indicates that for slower gradients (< 1.30 %Al/nm) where the bound polarization charge density is < 4×10$^{18}$ cm$^{-3}$, the negative bound polarization charge is partially compensated by deep-level donor states available near valence band leading to reduced free hole concentrations. For higher composition gradients at which the bound charge density is ≥ 4×10$^{18}$ cm$^{-3}$, the available deep-level donor states are saturated, and the free hole gas density is equal to the bound polarization charge density within the error of the capacitance measurements. Boosting the concentration gradient up to 4.95 %Al/nm provides a bound polarization charge density and closely equal hole concentration of 1.6×10$^{19}$ cm$^{-3}$.

It is well established that III-N surface acts as a reservoir for free electrons in polarization engineered devices [7], but not for free holes [8]. In planar 2D heterostructures with negative bound polarization charge, free holes are not observed due to the presence of deep-level hole traps in N-polar GaN/AlGaN thin films [17]. For the p-type graded nanowires studied here, such deep-level donor states are not present in high concentrations ($> 4\times10^{18}$ cm$^{-3}$), otherwise the measured capacitances for shorter graded base lengths would agree with the fully depleted model. In these polarization-engineered nanowire devices, the source of free holes can be ionized acceptors, acceptor-like surface states, and/or the highly doped p-type substrate. Although we have not measured the Mg (p-type dopant in AlGaN) concentration in these nanowires, we expect it to be very low (~$10^{17}$ cm$^{-3}$) due to the small sticking coefficient of Mg at the substrate temperatures used here. However, nanowires contain a very different population of surface states than epitaxial thin films. In c-plane planar devices the only exposed surface is either (0001) or (000$\bar{1}$), whereas in nanowires there is an m-plane {1$\bar{1}$00} coaxial sidewall in addition to the top (000$\bar{1}$) surface, which can facilitate the formation of acceptor-like states to provide free holes in the polarization doped p-graded region. For successful p-type polarization doping the total number of acceptor-like states has to be greater than the number of any compensating background donors. The ratio of acceptor-like states to background donors can be written as [11]

$$\frac{N_A}{N_D} = \frac{\sigma_A^{surf} \times 2\pi r h}{\rho_D^{back} \times \pi r^2 h} = \frac{2\sigma_A^{surf}}{\rho_D^{back} r}$$

where, $\sigma_A^{surf}$, $\rho_D^{back}$, $r$, and $h$ are surface acceptor density, background doping density, nanowire radius, and height of p-type section, respectively. Effective surface acceptor states in a-plane GaN is reported [18] to be ~$10^{12}$ cm$^{-2}$, which can provide, assuming $10^{17}$ cm$^{-3}$ background donor concentration [19], 10 acceptor atoms per background donor atom for a nanowire with 40 nm diameter leading to a hole density p~$N_A$. On the other hand, if the nanowire diameter were 400nm, the donor and acceptors would be equal and the nanowire fully compensated, p~0.

Therefore, to achieve a large acceptor to donor ratio, and thus maximize p, the nanowire diameter needs to be kept as small as possible. Moreover, in this study we grow nanowire heterostructures on highly doped p-type Si wafers, which can also provide free holes in the p-type section to compensate for negative polarization charge [20]. Full scale LED optical and electrical characterizations of these devices will be reported in a subsequent publication.

In conclusion, the p-type polarization doping in graded AlGaN nanowire light emitting diodes was systematically varied from 0.7 %Al/nm to 4.95 %Al/nm. Capacitance measurements reveal that for concentration gradients of less than 1.30 %Al/nm, the negative polarization-induced bound charge is partially compensated by both free holes and deep donors available near the valence band. For gradients greater than or equal to 1.30 %Al/nm, the deep donor concentration is negligible and free hole concentrations roughly equal to the bound polarization charge density are achieved up to $1.6 \times 10^{19}$ cm$^{-3}$ for concentration gradient of 4.95 %Al/nm.

Acknowledgements

This work was supported by the Army Research Office (W911NF-13-1-0329) and by the National Science Foundation CAREER award (DMR-1055164).

Table I. Nominal graded base length ($L_{nom}$), Achieved graded base length ($L_{base}$), Composition gradient (%Al/nm), and spontaneous polarization charge ($N_A^{pol}$) of the polarization graded devices.

| Device ID | $L_{nom}$ (nm) | $L_{grad}$ (nm) | %Al/nm | $N_A^{pol}$ (cm$^{-3}$) |
|---|---|---|---|---|
| P025 | 25 | 20.2 | 4.95 | 1.60891E+19 |
| P050 | 50 | 29.53 | 3.39 | 1.10058E+19 |
| P100 | 100 | 77.07 | 1.30 | 4.21695E+18 |
| P150 | 150 | 102.2 | 0.98 | 3.18004E+18 |
| P200 | 200 | 142.2 | 0.70 | 2.28551E+18 |

Table II. Plan view SEM fill factor ($ff_p$), base to top nanowire area ratio ($r_{bt}$), and quantum well to top area ratio ($r_{qt}$) for different devices.

| Device ID | $ff_p$ | $r_{bt}$ | $r_{qt}$ |
|---|---|---|---|
| P025 | 0.842 ± 0.030 | ~ 1.00 | ~ 1.00 |
| P050 | 0.822 ± 0.033 | 0.822 ± 0.094 | 0.971 ± 0.14 |
| P100 | 0.820 ± 0.041 | 0.596 ± 0.123 | 0.943 ± 0.05 |
| P150 | 0.823 ± 0.019 | 0.627 ± 0.114 | 0.976 ± 0.14 |
| P200 | 0.833 ± 0.033 | 0.694 ± 0.130 | 0.867 ± 0.07 |

Table III. Calculation of effective fill factor considering plan view SEM and tapered geometry.

| Device ID | $C_b$ (μF/cm$^2$) | $C_j$ (μF/cm$^2$) | $C_{cyl}$ (μF/cm$^2$) | $C_{tap}$ (μF/cm$^2$) | $ff_{tap}$ | $ff_{eff}$ |
|---|---|---|---|---|---|---|
| P025 | 0.426 | 0.267 | 0.164 | 0.164 | 1 | 0.842 |
| P050 | 0.421 | 0.249 | 0.156 | 0.142 | 0.910 | 0.748 |
| P100 | 0.382 | 0.207 | 0.134 | 0.105 | 0.783 | 0.642 |
| P150 | 0.368 | 0.196 | 0.128 | 0.105 | 0.818 | 0.673 |
| P200 | 0.350 | 0.181 | 0.119 | 0.095 | 0.799 | 0.666 |

Figure captions

Figure 1. (a) Schematic of nanowire heterostructure on p-type Si. (b) Strain distribution ($\varepsilon_{xx}$, $\varepsilon_{yy}$, $\varepsilon_{zz}$) in the graded nanowire heterostructure. Unit of length is nm and strain is shown in %. (c) Energy band diagram (top) and carrier concentration profile (bottom) along the axis of the nanowire heterostructure considering spontaneous and strain induced polarization charge.

Figure 2. (a) 1D circuit model for capacitance calculation with energy band diagram showing depletion regions corresponding to contributing capacitances. (b) Scanning transmission electron microscopy (STEM) images of graded nanowire heterojunctions. Nominal base lengths of the devices are shown at the bottom. Actual base lengths are extracted from STEM images for capacitance modeling. (c) Actual base length vs target (nominal) base length. Y-intersection of 16.8 nm represents the length of GaN section during nucleation. (d) Plan view SEM image analysis to find fill factor $ff_p$. (e) Schematic of a tapered nanowire showing $r_{qt}$ and $r_{bt}$.

Figure 3. The calculated capacitance from a simple 1D circuit model (triangles - Predicted with $h = N_A^{pol}$) and assuming full depletion (circles - Predicted with $h = 0$) are shown. The top horizontal axis corresponds to the negative bound spontaneous polarization charge density calculated from the measured concentration gradient. Squares are the extracted 1D nanowire capacitance from measured capacitance.

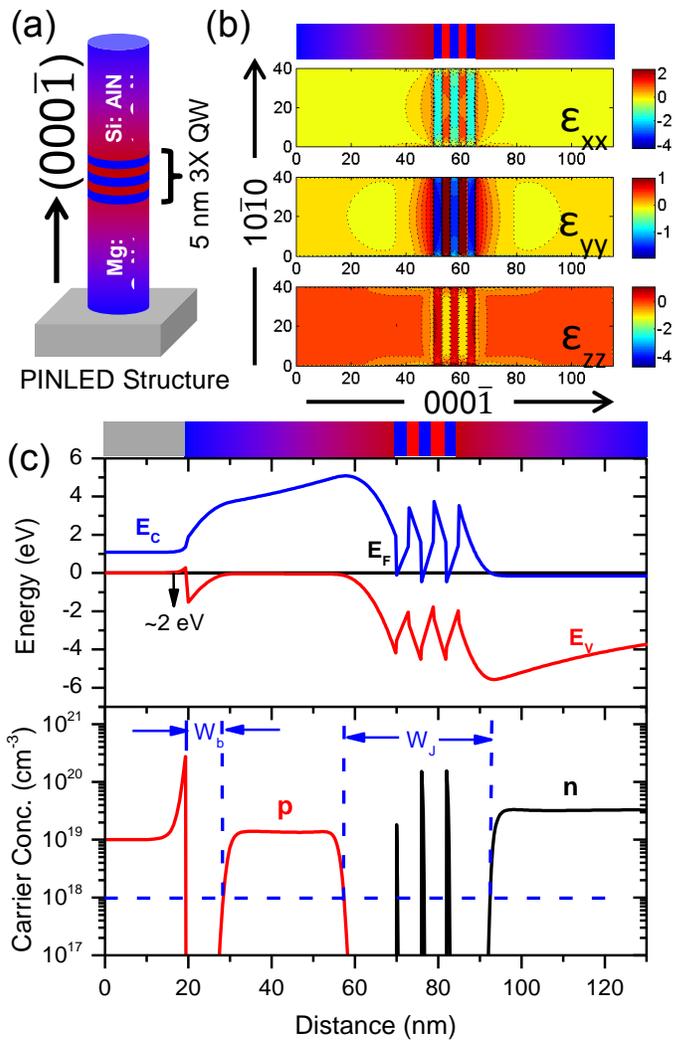

Figure 1

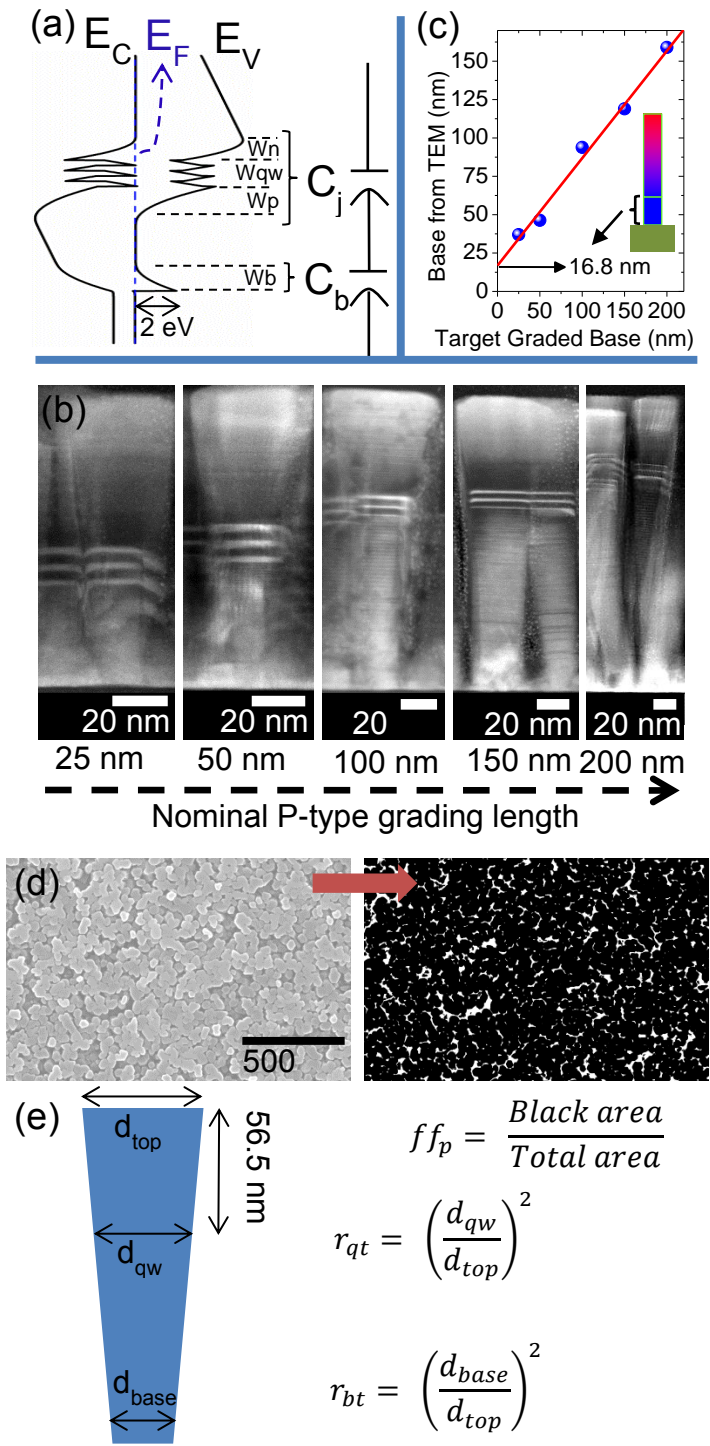

Figure 2

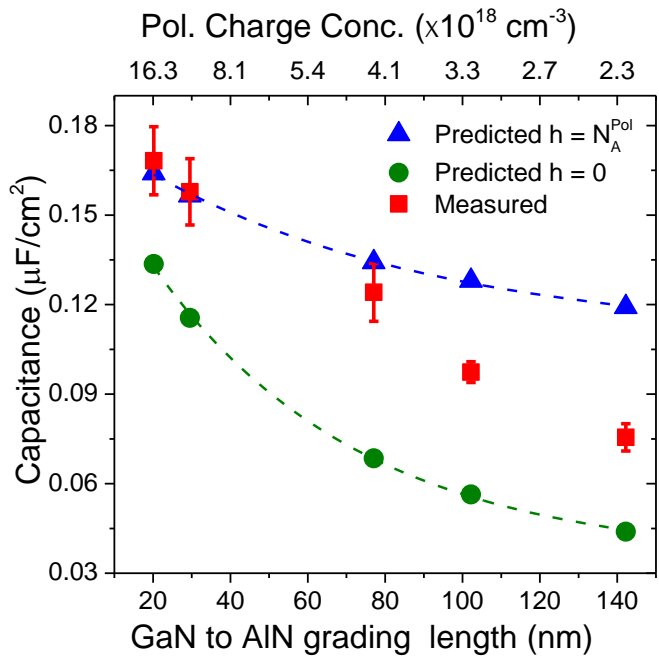

Figure 3